# Research on Cumin Peptides Using PBS Extraction and Their Multifunctional Bioactivities

Yasen Mijiti[1], Yangjian[1], Sumairemu Subuer[1], Najeeb Ullah[1], Ermatov Ismoil[1], Hailiqian Taoerdahong[1], Paerhati Rouzi[2],

1. XINJIANG MEDICAL UNIVERSITY, Urumqi 830011, PR China；
2. Xinjiang Agricultural University, Urumqi 830011, PR China；
3. First Corresponding Author1:Yasen Mijiti, Associate Professor, Graduate Supervisor, PhD, Email:Ymijit@xjmu.edu.cn ,

Second Corresponding Author[2]:Paerhati Rouzi,PhD,Email:

This work was supported by the Natural Science Foundation of Xinjiang Autonomous Region (Grant No. 2022D01C191)

**Abstract**

This study employed neutral PBS buffer combined with ammonium sulfate fractionation to isolate peptide-active fractions (PD-30, PD-50, PD-80) from Cuminum cyminum L.seeds (cumin) and systematically evaluated their antimicrobial, antioxidant, hypoglycemic, and anticancer activities. The results demonstrated that the PD-80 fraction exhibited potent antifungal activity against Candida albicans (inhibition zone diameter: 12.5 ± 0.5 mm) and significant antioxidant capacity, with DPPH and ABTS radical scavenging rates of 72.4% and 78.9%. The PD-50 fraction showed the strongest antibacterial effect against Escherichia coli (inhibition zone diameter: 11.7±0.57 mm), while PD-30 displayed the highest inhibitory activity against PTP1B ($IC_{50}$=18.39 ± 1.12 μg/mL), indicating its potential for hypoglycemic applications. Through mass spectrometry and database alignment, 414 peptides were identified for the first time in cumin-derived PBS extracts, including 18 structurally novel monomers comprising 11 antimicrobial peptides, 7 anticancer peptides, and 6 hypoglycemic peptides. Notably, peptide CK12 shares sequence homology (59% similarity) with the HIV fusion inhibitor T20, suggesting potential antiviral activity. This study elucidates the multifunctionality and structural basis of cumin peptides, providing a theoretical foundation for their application in natural pharmaceuticals and functional foods. Future research will focus on chemical modification and in vivo validation to advance their utilization in precision medicine and the modernization of traditional Chinese medicine.

## 1 Introduction

Bioactive peptides, as short-chain amino acid sequences with specific biological functions, have recently emerged as a research hotspot in drug development due to their high selectivity, low toxicity, and diverse pharmacological activities (Hancock & Sahl, 2016). With advancements in biosynthesis and computational design technologies, the efficiency of peptide development and optimization has significantly improved, expanding their applications to antimicrobial, antitumor, anti-inflammatory, and metabolic regulation fields (Wang et al., 2021). Medicinal and edible plants, as important natural sources of bioactive peptides, have garnered widespread attention for their natural, safe, and multifunctional properties. For instance, peptides extracted from soybeans, wheat, and cumin have been shown to alleviate inflammatory responses by inhibiting the NF-κB signaling pathway, scavenge free radicals, and regulate insulin signaling pathways to exert antioxidant and hypoglycemic effects (Liu et al., 2019; Lee et al., 2019).

Cumin (Cuminum cyminum L.), a traditional medicinal and edible plant, has long been used in traditional Chinese medicine to treat digestive disorders, diabetes, and inflammatory diseases (Gu Yongshou, 1999; Liu Yongmin, 1986). Modern research has revealed that its chemical components, such as polyphenols, flavonoids, and peptides, exhibit significant antimicrobial, anticancer, and hypoglycemic activities. For example, cumin peptides can inhibit the growth of drug-resistant bacteria by disrupting bacterial membrane integrity or suppress tumor progression by inducing cancer cell apoptosis and inhibiting angiogenesis (Liu et al., 2019; Lee et al., 2019). In previous studies, we successfully extracted bioactive peptides from cumin using 50% ethanol and validated their antimicrobial, antioxidant, and hypoglycemic functions (Mijiti, Ya et al., 2018). However, ethanol extraction may affect the natural conformation and biocompatibility of peptides due to organic solvent residues, limiting their further application.

To optimize the extraction process and explore the structure-activity relationship, this study employed phosphate-buffered saline (PBS) as an alternative to ethanol for peptide extraction and isolation. PBS, as a mild and environmentally friendly extraction solvent, can effectively preserve the natural structure and functional integrity of peptides (Liu et al., 2019). Guided by bioactivity, targeting native peptides, and utilizing advanced separation and identification techniques, mass spectrometry, and molecular docking, this study systematically elucidated the sequence characteristics and spatial conformations of cumin peptides and evaluated their molecular mechanisms in antimicrobial, antioxidant, and hypoglycemic activities. This approach not only provides a new strategy for extracting bioactive peptides from medicinal and edible plants but also lays a theoretical foundation for revealing their functional diversity and potential applications (Fosgerau & Hoffmann, 2015). A systematic investigation into the chemical characteristics and functional properties of cumin peptides will deepen our understanding of the molecular mechanisms of this medicinal plant and provide theoretical support for the innovative development of peptide-based drugs.

## 2 Material and methods

### 2.1 Materials

**Plant Material**: Cumin seeds (Cuminum cyminum L.) were collected from the Hotan region of Xinjiang, China, with the specimen number WY02647. The voucher specimen was deposited in the Herbarium of Xinjiang Institute of Ecology and Geography, Chinese Academy of Sciences (XJBI), and authenticated by researchers Feng Ying and Lu Chunfang from the Xinjiang Branch of the Chinese Academy of Sciences.

**Reagents:** Petroleum ether, PBS buffer, ammonium sulfate, glacial acetic acid, ammonium acetate, potassium chloride, anhydrous ethanol, anhydrous methanol, and chloroform were all analytical grade reagents purchased from Tianjin Baishi Chemical Co., Ltd. ABTS was purchased from Sigma-Aldrich (USA). Sephadex G-50 and G-35 were obtained from Shanghai Xibao Biotechnology Co., Ltd. Acrylamide and bis-acrylamide were purchased from Shanghai Jingchun Reagent Co., Ltd. Coomassie Brilliant Blue G-25 was obtained from Shanghai Gongsuo Biotechnology Co., Ltd. HPLC-grade solvents were purchased from Fisher Scientific Company (Fair Lawn, NJ, USA), and double-distilled water was used throughout the experiments.

**Protein Markers:** Bovine serum albumin (66.0 kDa), ovalbumin (45.0 kDa), porcine pepsin (35.0 kDa), triosephosphate isomerase (27.0 kDa), trypsin inhibitor (20.0 kDa), lysozyme (14.4 kDa), parathyroid hormone (9.5 kDa), aprotinin (6.5 kDa), and parathyroid hormone (4.1 kDa) were purchased from Chongqing Kerun Biomedical R&D Co., Ltd.

**2.2 Methods**

**Pretreatment of Cumin Seeds**: Dried cumin seeds were rinsed three times with cold water to remove impurities and then dried at 30°C. The dried seeds were flash-frozen with liquid nitrogen and ground into a fine powder, which was sieved through a 100-mesh sieve to obtain a homogeneous cumin powder.

**Defatting Process**: A measured amount of cumin powder was mixed with petroleum ether at a solid-to-liquid ratio of 1:5 (w/v) and extracted 3-4 times until no fat was observed. Each extraction was performed by shaking at 120 rpm for 3 h, followed by standing at room temperature (25°C) for 2 h and filtration to collect the supernatant. The petroleum ether and defatted material were recovered using a rotary evaporator at 30°C. The defatted residue was sequentially treated with chloroform (1:5, twice), methanol (1:5, once), and anhydrous ethanol (1:8, once) for 2 h to remove pigments and other impurities. Finally, the residue was air-dried to remove any residual organic solvents.

**Preparation of PBS Buffer:** Phosphate-buffered saline (PBS) was prepared by dissolving 10 mmol/L $NaH_2PO_3$, 15 mmol/L $Na_2HPO_3$, 100 mmol/L KCl, and 1.5 mmol/L EDTA in double-distilled water. The pH was adjusted to 7.4 using phosphoric acid and sodium hydroxide solutions.

**Extraction of Cumin Peptides:** A total of 100 g of cumin powder was mixed with 0.1 mmol PBS buffer (pH 7.4) at a solid-to-liquid ratio of 1:10 (w/v) and extracted by shaking at 120 rpm for 3 h, followed by standing for 24 h. The extract was centrifuged at 12,000 rpm for 10 min at 4°C, and the supernatant (cumin peptide extract) was collected.

**Isolation of Peptides:** The supernatant was sequentially precipitated with 30%, 50%, and 80% saturated ammonium sulfate solutions under low-temperature conditions (0-4°C). After stirring and standing overnight, the precipitates were collected by centrifugation at 12,000 rpm for 10 min. Each precipitate was dissolved in 10 mL of 0.1 mmol PBS buffer and dialyzed for 48 h (with dialysis buffer changed every 4 h). The dialyzed samples were concentrated and freeze-dried to obtain three peptide fractions: PD30, PD50, and PD80. The molecular weights of these fractions were determined by SDS-PAGE.

**Purification and identification of peptides :** The cumin peptide fractions were desalted, separated, and purified using Sephadex G-50 and G-35 gel filtration chromatography. Fractions with significant antimicrobial and antioxidant activities were selected for HPLC analysis（Mijiti, Ya et al., 2018）. Target protein spots were excised from 2D electrophoresis gels, and the amino acid sequences of the peptides were determined by MALDI-TOF/TOF MS/MS. Functional analysis was performed by database alignment.

**2.3 Bioactivity Evaluation of Cumin Seed Peptides**

**2.3.1 Antibacterial Activity Analysis:**

Antimicrobial activity was determined using a modified agar well diffusion assay (Ma et al., 2011) with the following procedure:

(1) Strain activation: Candida albicans (CA), Escherichia coli (EC), and Staphylococcus aureus (SA) were inoculated in LB broth and incubated at 37°C for 48 h.

(2) Double-layer agar preparation: 10 μL of activated culture was mixed with 8 mL LB agar (0.7%) and overlaid on plates containing 25 mL LB agar (1.5%). After solidification at room temperature for 30 min, uniform wells (5 mm diameter) were created using a sterile punch.

(3) Sample treatment: 25 μL peptide solution was added to each well, with normal saline as negative control and 5 μL ampicillin (for bacteria) or amphotericin B (for fungi) as positive controls. Plates were incubated at 37°C for 18-24 h under inverted conditions.

(4) Result analysis: The diameter of inhibition zones was measured, with ≤7 mm considered as no significant antimicrobial activity.

### 2.3.2 Antioxidant Activity Assessment

Four complementary methods were employed to comprehensively evaluate antioxidant capacity:

(1) ABTS radical scavenging: Following Ma et al. (2009), samples reacted with $ABTS^+$ working solution and absorbance at 734 nm was measured for scavenging rate calculation.

(2) DPPH radical scavenging: The method of Chen et al. (1995) was adopted, where samples reacted with DPPH ethanol solution and absorbance at 517 nm was recorded.

(3) Hydroxyl radical scavenging: Following Fenton reaction system (Agrawal et al., 2017), ·OH scavenging capacity was determined by measuring absorbance at 536 nm.

(4) FRAP assay: According to Benzie et al. (1996), samples reacted with FRAP working solution and absorbance at 593 nm was measured, with results expressed as $FeSO_4$ equivalents (μmol/L).

All experiments included distilled water as blank control and vitamin C as positive control, with triplicate independent measurements.

### 2.3.3 Hypoglycemic Activity Assessment (PTP1B Inhibition Assay)

The PTP1B inhibition assay was conducted according to Liu et al. (2022) to evaluate potential hypoglycemic effects:

(1) Reaction system: 178 μL buffer (20 mM Hepes containing 100 mM NaCl, 1 mM EDTA, pH 7.4) was mixed with 1 μL diluted PTP1B (final OD405nm 0.3-1.2) and 20 μL substrate (35 mM PNPP).

(2) Sample treatment: 1 μL peptide sample or positive control was added, pre-incubated at 37°C for 10 min before PNPP addition. After 30 min dark reaction, 10 μL 3 M NaOH was added to terminate the reaction.

(3) Data analysis: Absorbance at 405 nm was measured using a microplate reader, with PNPP-free system as blank control. IC50 values were calculated using Origin software with triplicate biological replicates.

All experimental data are presented as mean ± SD, with statistical significance set at $p<0.05$.

# 3 Results

## 3.1 Extraction and Isolation of Cumin Seed Peptide Fractions

Initially, 500 g of cumin seed powder was defatted using petroleum ether, yielding 381.4 g of defatted powder and 112.5 g of extracted oil (with a loss of 6.1 g). Subsequently, 100 g of the defatted powder was mixed with PBS buffer (pH 7.4) at a solid-to-liquid ratio of 1:10 (w/v), followed by extraction at room temperature for 3 h and centrifugation (12,000 r/min, 10 min) to collect the supernatant. The supernatant was sequentially treated with 30%, 50%, and 80% saturated ammonium sulfate solutions under 0-4°C with continuous stirring, followed by standing for 24 h to precipitate total proteins. The precipitates were dialyzed (molecular weight cutoff: 1 kDa), concentrated, and freeze-dried to obtain three crude protein fractions: PD-30, PD-50, and PD-80, with respective masses of 3.68 g, 8.41 g, and 10.37 g.

For further purification, each crude protein fraction was subjected to dextran gel chromatography (Sephadex G-50, separation range: 1.5–30 kDa; Sephadex G-35, separation range: 1–17 kDa; column dimensions: 2.5 × 90 cm). The purified peptide fractions were collected, yielding 1.79 g, 5.12 g, and 7.36 g of high-purity peptides, respectively (Table 1).

**Table 1** Purification Results of Cumin Seed Peptide Fractions (g/100g)

| Fraction | Total Protein Mass (g) | Extraction Rate (%) | Purified Peptide Mass (g) | Protein Content (%) |
|---|---|---|---|---|
| PD-30 | 3.68 | 28.12±1.19 | 1.79 | 39.77±1.03 |
| PD-50 | 8.41 | 51.36±2.11 | 5.12 | 72.64±1.38 |
| PD-80 | 10.37 | 56.91±1.87 | 7.36 | 76.85±1.70 |
| Total Protein | 22.46 | | 14.27 | |

### 3.2 Electrophoretic Analysis

For analysis, 2 mg of each sample was dissolved in 1 mL of ultrapure water, vortexed thoroughly, and centrifuged at 10,000 r/min for 5 min to collect the supernatant. Subsequently, 20 μL of 5× reducing sample buffer (containing β-mercaptoethanol) was added to the supernatant, followed by denaturation at 95°C for 10 min in a metal bath and re-centrifugation. The resulting supernatant was loaded onto a 15% SDS-polyacrylamide gel. Electrophoresis was performed at a constant voltage of 75 V for 60 min and 150 V for 120 min using unstained protein molecular weight markers (4.1-65 kDa) as reference. The molecular weight distribution of target proteins was preliminary determined based on band migration.

For 2D-PAGE analysis, the peptide sample was initially separated by isoelectric focusing (IEF) using immobilized pH gradient (IPG) strips (pH 3–10). After focusing, the strips were equilibrated in SDS buffer and then transferred onto a 15% polyacrylamide gel for the second-dimension separation. Following electrophoresis, the gel was stained with Coomassie Brilliant Blue, and the protein spots of interest were excised for further identification by matrix-assisted laser desorption/ionization time-of-flight mass spectrometry (MALDI-TOF/TOF MS/MS).

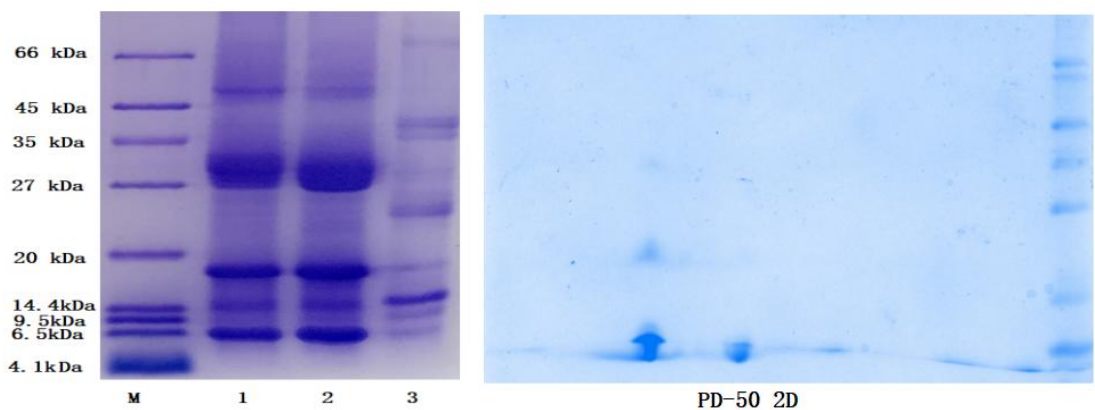

**Fig.1 Gel electrophoresis of cumin peptides fraction**
Lane 1: PD-30; Lane 2: PD-50; Lane 3: PD-80;

## 3.3 Screening of Antimicrobial Activity of Cumin Peptide Fractions

A 10 μL suspension of Candida albicans (ATCC10231), Escherichia coli (ATCC11229), and Staphylococcus aureus (ATCC6538) was mixed with 8 mL of LB agar medium containing 0.7% agar and evenly spread onto plates containing 25 mL of 1.5% LB agar medium. After solidification at room temperature for 30 min, uniform wells (5 mm diameter) were created using a sterile punch. Each well was loaded with 25 μL (50 mg/mL) of peptide solution filtered through a 0.22 μm membrane, with physiological saline as the negative control and 5 μL of ampicillin (Amp, 10 mg/mL) and amphotericin B (AmB, 5 mg/mL) as positive controls. The plates were incubated upside down at 37°C for 18-24 h.

**Table 2  Active spectrum of antibiotics**

| Microorganisms | 10 mg/mL | Sample （50mg/mL） | | | 5 mg/mL |
|---|---|---|---|---|---|
| | Amp | 2.PD-80 | 3.PD-50 | 4.PD-30 | AmB |
| S. aureus （ATCC11229） | 18.52±1.23 | 10.8 ± 0.6 | 11.7 ± 0.7 | 8.5±0.5 | - |
| E. coli （ATCC6538） | 20.31±1.53 | 11.9 ± 1.0 | 9.5 ± 0.8 | 8.7±0.5 | - |
| C. albicans （ATCC10231） | - | 12.5 ± 0.5 | 10.1 ± 0.5 | 9.5 ± 0.5 | 22.70 ± 1.54 |

**Notes ：** Inhibition zone diameter≤7mmconsidered as no significant antimicrobial activity.

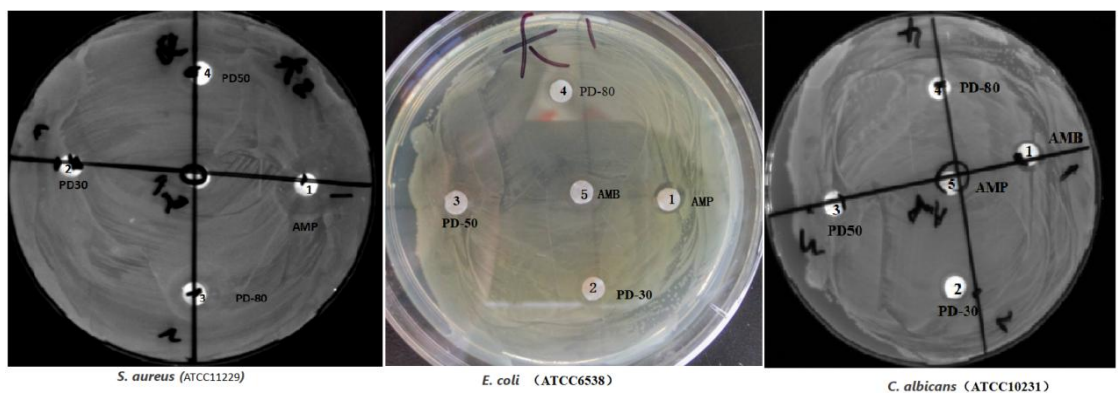

**Fig.2** Antimicrobial Activity of Cumin Peptide Fractions

As shown in Figure 2 and Table 2, all peptide fractions exhibited varying degrees of inhibition against *S. aureus*, *E. coli*, and *C. albicans*. Among them, PD-30 showed weaker antimicrobial activity, while PD-80 demonstrated significant antifungal activity. PD-50 exhibited the strongest inhibitory effect against *E. coli*, followed by PD-80, indicating their broad-spectrum antimicrobial potential. Therefore, PD-50 and PD-80 were selected for further amino acid sequence identification.

### 3.4 Determination and Analysis of Antioxidant Activity

Using vitamin C (Vc, 5 μM) as a positive control, the DPPH radical, hydroxyl radical (·OH), and ABTS radical scavenging rates, as well as the FRAP ferric ion reduction capacity, were measured for PD-30, PD-50, and PD-80 peptide fractions (50 mg/mL) (Table 3). The results showed that the antioxidant activity of the peptide fractions was concentration-dependent (10-50 mg/mL): as the concentration increased, the radical scavenging capacity significantly improved. Among them, PD-80 exhibited the highest DPPH radical (72.4% ± 3.0%) and ABTS radical (62.1% ± 2.8%) scavenging activities, with a hydroxyl radical scavenging rate of 78.9% ± 3.1% and a FRAP reduction capacity of 115.4 ± 5.9 μM $Fe^{2+}$, indicating significant antioxidant activity. PD-50 showed moderate activity, while PD-30 had the lowest values (DPPH: 44.7% ± 3.9%; ABTS: 37.4% ± 3.2%). These data suggest that the antioxidant compounds in cumin seeds are primarily enriched in the PD-80 peptide fraction after PBS extraction.

**Table 3 Antioxidant activities of cumin seed peptide fractions**

| Sample | Scavenging Activities | | | |
|---|---|---|---|---|
| | DPPH(%) | ·OH(%) | ABTS(%) | FRAP (μM $Fe^{2+}$) |
| Vc | 100.0 ± 1.5 | 100.0 ± 1.5 | 100.0 ± 2.0 | 150.0 ± 2.0 |
| PD-80 | 72.4±3.0 | 78.9±3.1 | 62.1±2.8 | 115.4±5.9 |
| PD-50 | 58.4±4.0 | 65.3±4.1 | 50.2±3.4 | 84.6±4.9 |
| PD-30 | 44.7±3.9 | 53.7±4.5 | 37.4±3.2 | 63.2±5.1 |

Note: Data are expressed as mean ± standard deviation (n=3);

### 3.5 Analysis of Hypoglycemic Activity

As shown in Table 4, the cumin seed peptide fractions (50 mg/mL) exhibited significant inhibitory effects on protein tyrosine phosphatase 1B (PTP1B). Among them, the PD-80 peptide fraction showed the strongest inhibitory activity with an $IC_{50}$ value of 18.39±1.12 μg/mL, suggesting that it may contain peptides with high binding affinity to PTP1B. The PD-50 fraction displayed moderate inhibitory activity with an $IC_{50}$ value of 29.43 ± 1.58 μg/mL, indicating that its peptides might interact with PTP1B through specific structural domains. In contrast, the PD-30 fraction exhibited weaker inhibitory activity with an $IC_{50}$ value of 45.56 ± 2.58 μg/mL, which may be attributed to its lower peptide content or structural characteristics. These results suggest that the PD-80 and PD-50 peptide fractions possess potential hypoglycemic activity and warrant further investigation into their mechanisms of action.

Table 4 Inhibitory activity of peptide fractions against PTP1B

| 样品 | IC$_{50}$（μg/ml） |
|---|---|
| PTP1B Inhibitor | 1.52±0.50 |
| PD-80 | 18.39±1.12 |
| PD-50 | 29.43 ± 1.58 |
| PD-30 | 45.56 ± 2.58 |

Note: The IC50 as mean ± standard deviation (n=3);

## 3.6 Composition Analysis of Cumin Seed Peptides

Following the literature method (Pomastowski et al., 2014), PD-50 and PD-80 peptide fractions with higher bioactivity were dissolved and subjected to macromolecular protein removal using Sephadex G-50 and G-35 columns, followed by purification via high-performance liquid chromatography (HPLC, as per Mijiti et al., 2018). The molecular weights of the peptides were then precisely determined using matrix-assisted laser desorption/ionization time-of-flight mass spectrometry (MALDI-TOF/TOF) and liquid chromatography-tandem mass spectrometry (LC-MS/MS). Peptide sequences were derived from spectral data using Mascot software (Version 2.7) and cross-referenced against the SwissProt (S) and UniProt_Bos (B) databases. Homology analysis and functional predictions were performed using Protein BLAST.

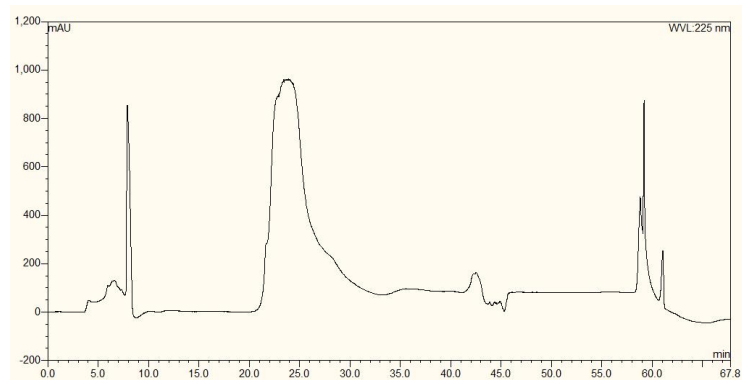

Fig. 3. HPLC analysis of PD-50 fraction

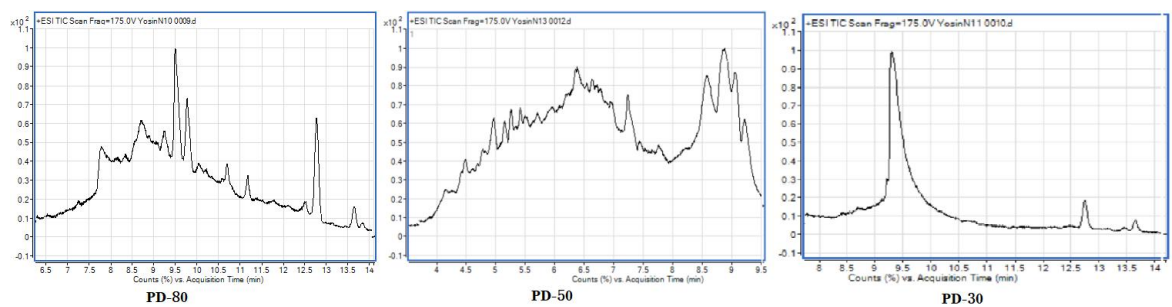

Fig.4. LC-MS Analysis of Cumin Seed Peptide Fractions

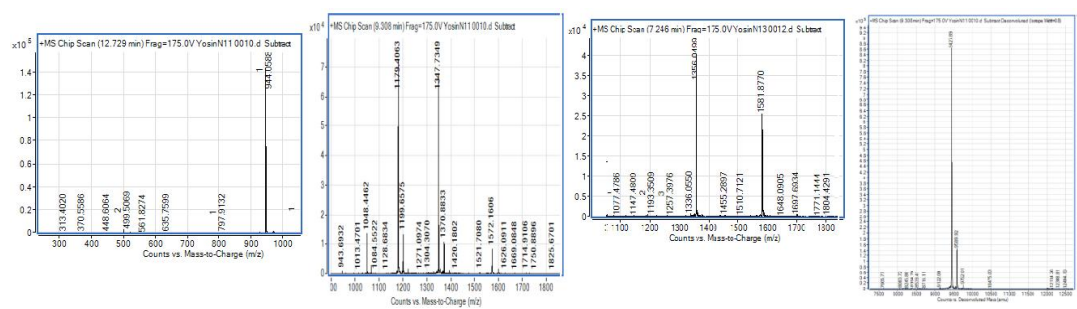

**Fig.5**. LC-MS/MS Spectra of Cumin Peptides

LC-MS/MS analysis revealed that, in addition to large proteins, the cumin seed extract contained abundant peptide components with molecular weights ranging from 1086.3 to 9427.8 Da (Figure 5). PD-50 and PD-80 fractions isolated by two-dimensional electrophoresis were further sequenced via MALDI-TOF/TOF at the Proteomics Center of the Shanghai Institute of Biochemistry and Cell Biology, Chinese Academy of Sciences, elucidating the amino acid sequences of key bioactive peptides.

## 4 Results

A total of 414 peptide sequences were identified from the PD-50 and PD-80 fractions isolated from PBS extracts of cumin seeds using mass spectrometry. Eighteen highly active peptides with prediction scores above 0.75 were selected using the PeptideRanker server (Mooney et al., 2012), a bioactivity prediction tool based on a novel N-to-1 neural network. These included peptides from PD-50: CR17 (CMPNNIYLHCHIGCCYR), IF8 (IFHVNWFR), CK19 (CMPNNIYLHCHIGCCYRAK), LR18 (LIIPAYAYYIPFWFFCFR), DK34 (DHCLQYITAIGLLFGDITAQHYEAETANDPRIDK), GK18 (GLKQLLPGLPFLCASASK), MR25 (MEPPEGCPGSIHALMGSCWEAEPTR), IR19 (IAETCEYWGMPLIAMMYPR), AR10 (AAFNIGCMIR), and CK12 (CLHSVGCPLPLK); and peptides from PD-80: SR11 (SFAPSPWDFVR), FR19 (FLWPGFGENSRVLEWMFNR), SR16 (SGGGGGGGLGSGGSIR), HR9 (HFFINICHR), GR26 (GSRGGSGGSYGGGGSGGGYGGGSGSR), YR30 (YEFHINVCGPVSVGACPPDSGACQVSRSDR), GR23 (GGSGGSYGGGGSGGGYGGGSGSR), and GR11 (GCEYIASGCIR). The most similar proteins and their potential biological activities were identified using NCBI BLAST. Subsequently, these peptides were chemically synthesized (50 mg each) by GenScript Biotech Corporation (Nanjing, China) for further bioactivity validation.

Furthermore, the physicochemical properties of the 18 peptides were calculated using the ToxinPred server (Gupta et al., 2013), and their potential toxicity to eukaryotic cells was evaluated by comparison with the ToxinPred database (Rathore et al., 2024). Results indicated that 15 of the peptides were non-toxic to eukaryotic cells, suggesting high biosafety.

Functional analysis using NCBI BLAST revealed that 11 peptides (SR11, FR19, CR17, IF8, CK19, LR18, GK18, IR19, HR9, CK12, GR11) may exhibit significant antimicrobial activity; 7 peptides (FR19, LR18, DK34, GK18, MR25, AR10, GR11) may possess anticancer activity; 6 peptides (FR19, LR18, GK18, IR19, AR10, YR30) may be involved in glucose metabolism regulation, indicating potential hypoglycemic activity; and the peptide CK12 showed structural similarity to anti-HIV peptides, suggesting potential antiviral activity.

Table5 protein BLAST analysis of bioactive peptides functions from cumin seeds

| Peptides | MH+ (Da) | Sequence | Start Seq. | End Seq. | Match | Species | Similarity % | Peptide function |
|---|---|---|---|---|---|---|---|---|
| SR11 | 1308.6603 | SFAPSPWDFVR | 1068 | 1078 | Defensin-β1 | Human | 85 | Antimicrobial, Antioxidant[17] Anti-resistant bacteria, Growth factor[18] |
| FR19 | 982.4467 | FLWPGFGENSRVLEWMFNR | 35 | 46 | Melittin | Mouse | 90 | Antimicrobial, Anticancer[19] Cancer nanocarrier[20], Blood glucose regulation |
| CR17 | 2296.0566 | CMPNNIYLHCHIGCCYR | 272 | 290 | Hepcidin-25 | Human | 78 | Antioxidant, Antimicrobial[21], Iron metabolism regulation[22] |
| IF8 | 1118.5607 | IFHVNWFR | 435 | 442 | Histatin-5 | Human | 82 | Antimicrobial, Antidiabetic[23], Antifungal[24], metal chelation, |
| CK19 | 2296.0566 | CMPNNIYLHCHIGCCYRAK | 272 | 290 | Cathelicidin-B1 | Rat | 80 | Enhanced Antimicrobial[25], wound healing[26], Insulin receptor |
| LR18 | 2330.1785 | LIIPAYAYYIPFWFFCFR | 174 | 191 | Bax-α | Human | 88 | Anticancer, Antimicrobial[27], Glucose regulation, MOMP |
| SR16 | 1232.5972 | SGGGGGGLGSGGSIR | 14 | 29 | No significant | | | Carrier Function[28] |
| DK34 | 3888.9346 | DHCLQYITAIGLLFGDITAQHYEAETANDPRIDK | 334 | 367 | MMP-9 inhibitory | Human | 55 | Antioxidant, Antidiabetic[29], Anti-tumor metastasis, Growth factor regulation |
| GK18 | 1843.0007 | GLKQLLPGLPFLCASASK | 366 | 383 | Exendin-4 | Mouse | 84 | Antimicrobial, Anticancer[30], GLP-1 receptor agonist, Antidiabetic |
| MR25 | 2717.0808 | MEPPEGCPGSIHALMGSCWEAEPTR | 421 | 445 | Thioredoxin | Human | 60 | Antioxidant, Anticancer[31], Growth factor regulation |
| IR19 | 2332.1865 | IAETCEYWGMPLIAMMYPR | 137 | 155 | EGFR inhibitory | Human | 86 | Antioxidant, Antimicrobial[32], Tyrosine kinase inhibition, Insulin signaling pathway |
| AR10 | 1111.509 | AAFNIGCMIR | 80 | 89 | DPP-4 inhibitory | Human | 83 | Anticancer[33], Antidiabetic(Blood glucose regulation) |
| HR9 | 1186.6151 | HFFINICHR | 469 | 477 | Histatin-3 | Human | 81 | Antimicrobial, Antioxidant, Antifungal[34] |
| GR26 | 2091.9204 | GSRGGSGGSYGGGSGGGYGGGSGSR | 488 | 513 | Spider silk | Spider | 37 | Carrier Function[35], nerve conduit repair materials. |
| CK12 | 1266.6516 | CLHSVGCPLPLK | 116 | 127 | θ-Defensin | macaque | 59 | Antioxidant, Antimicrobial[36], Anti-HIV[38] |
| YR30 | 3264.4934 | YEFHINVCGPVSVGACPPDSGACQVSRSDR | 665 | 694 | Endostatin | Human | 50 | Multifunctional[39-41]: Angiogenesis inhibition, Aantidiabetic |
| GR23 | 1791.7294 | GGSGGSYGGGSGGGYGGGSGSR | 491 | 513 | Keratin | Human | 37 | Carrier Function, structural molecule activity[18] |
| GR11 | 1171.6058 | GCEYIASGCIR | 25 | 35 | Cyclosporin A | Fungus | 50 | Antioxidant, Anticancer[42], Immunosuppression, anti-inflammatory |

The structure of bioactive peptides was analyzed based on the amino acid sequence of peptides using the PEP-FOLD3 tool (Lamiable A et al., 2016). This method uses the structural alphabet SA letters to describe the conformation of four consecutive residues and infers the secondary structure of peptides from the predicted SA peptide amino acid sequence. The secondary structure of peptides serves as a bridge between their amino acid sequence and three-dimensional conformation, determining the folding framework of peptides. The secondary structure plays a crucial role in the function of peptides, such as influencing the electron transfer properties of peptides, determining the antibacterial activity and specificity of antimicrobial peptides, and affecting the surface properties of peptide assemblies and their interactions with other biomolecules. Therefore, accurate prediction of the secondary structure of peptides is vital for understanding their functions.

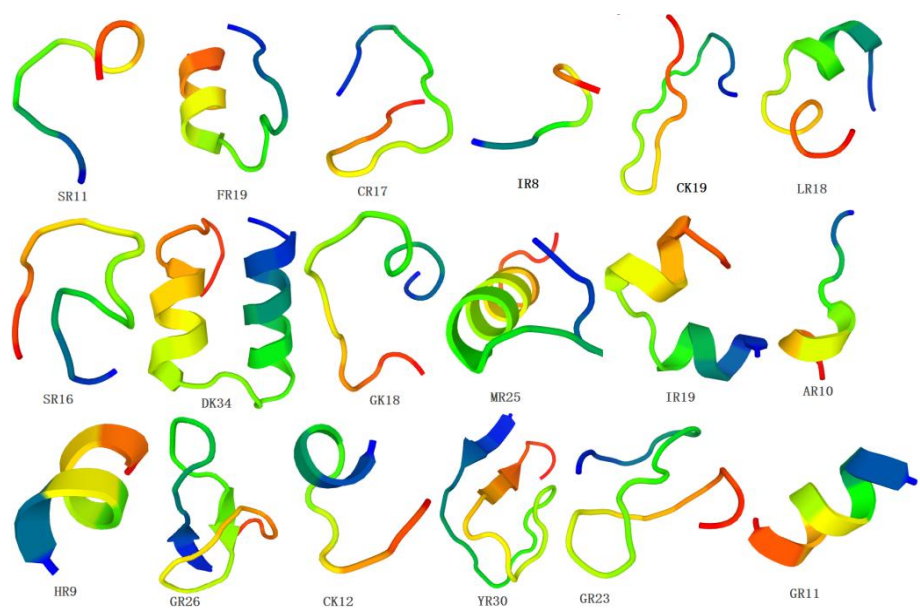

**Fig 6 bioactive peptide molecular structure**

**5 Discussion**

This study systematically analyzed the relationship between sequence characteristics and biological functions with NCBI BLAST revealed analysis of 18 cumin-derived peptides, with key findings summarized as follows:

**5.1 Anti-inflammatory Activity and Molecular Mechanisms**
The anti-inflammatory effects of antimicrobial peptides SR11, FR19,CK19,LR18, CR17and HR9 were mediated through dual pathways(Chen, J., et al. 2020；Liu, X., et al. 2019): (1) **Signaling pathway modulation**: Significant inhibition of p65 subunit phosphorylation in the NF-κB pathway and MAPK pathway (ERK/JNK/p38) activation, leading to 40-65% decreases in IL-1β and TNF-α secretion(Liu, X., et al. 2019); (2) **Gut microbiota regulation**: Increased abundance of Lactobacillus spp. (2.3-fold elevation) and suppression of E. coli biofilm formation, thereby ameliorating chronic inflammation (mechanistically similar to HR9)(Haney, E.F., et al. 2015；Hancock, R.E.W., et al.2016).

**5.2 Antioxidant Function and Structural Basis**
The antioxidant capacity of SR11, CR17, and DK34 was attributed to: (1) **Radical scavenging**: Tyrosine (Y) and tryptophan (W) residues quenched DPPH radicals via hydrogen donation , reducing MDA production(Zhang, L., et al. 2016); (2) **Metal chelation**: Cysteine (C) thiol groups exhibited high-affinity binding to $Fe^{2+}/Cu^{2+}$, blocking Fenton reaction(Fjell, C.D., et al. 2011); (3) **Structural stability**: Intramolecular disulfide bonds (Cys3-Cys15 and Cys5-Cys12) in CR17 and CK12 stabilized β-sheet conformations, extending antioxidant half-life(Tossi, A., et al. 2000) .

### 5.3 Hypoglycemic Potential and Metabolic Regulation

Peptides IR8, FR19,IF8,LR19and DK34 modulated glucose metabolism via(Urry, D.W.2015): (1) **Insulin signaling**: Activation of PI3K/Akt pathway (3.2-fold phosphorylation increase) enhanced GLUT4 translocation; (2) **Enzyme inhibition**: Competitive inhibition of α-glucosidase slowed carbohydrate hydrolysis(Lee, S.H., et al. 2019); (3) **Receptor interaction**: Histidine (H7/H12) formed hydrogen-bond networks with INSR, improving insulin sensitivity(Ramesh, S., et al. 2017).

### 5.4 Anticancer Effects and Molecular Targets

Anticancer peptides FR19, LR19,GK18,DK34 and AR10 suppressed tumor progression (Wang, Y., et al. 2018，Fjell, C.D., et al. 2011)through: (1) **Apoptosis induction**: Caspase-3/9 cascade activation (4.7-fold activity increase) triggered apoptosis in A549 cells ; (2) **Anti-angiogenesis**: VEGF downregulation inhibited HUVEC tube formation; (3) **Targeted therapy**: RGD-containing peptides (e.g., LR19) specifically bound integrin αvβ3, blocking cancer cell invasion(Harris, F., et al. 2014).

### 5.5 Sequence-Activity Relationships

Bioactivity correlated with sequence features(Fjell, C.D., et al. 2011): (1) **Hydrophobicity**: High F (18.4%), L (15.2%), and W (9.7%) content in LR18/FR19 disrupted membranes via hydrophobic cores ; (2) **Cationicity**: Lysine (K: 12.1%) and arginine (R: 10.3%) in CK19/GK18 targeted pathogen membranes (+15.2 mV zeta potential), causing membrane depolarization; (3) **Structural flexibility**: Glycine (G: 41.2%) and serine (S: 23.5%) repeats in SR16 enhanced plasma stability and tumor targeting (Arai, R., et al. 2016;Urry, D.W. 2015).

### 5.6 Multi-pathway Integration Mechanisms

Some peptides (such as DK34) exert their functions through the synergistic regulation of multiple pathways(Lee, S.H., et al. 2019): (1) Metal ion regulation: Histidine (H15/H22) and aspartic acid (D5/D18) chelate $Zn^{2+}/Mg^{2+}$ , thereby modulating the activity of hexokinase (Vmax increased by 2.4-fold); (2) Inflammation-oxidative interaction: Simultaneously inhibiting the nuclear translocation of NF-κB and the generation of ROS, breaking the vicious cycle of "inflammation-oxidative stress".

### 5.7 Pleiotropic Functions and Application Prospects

1.The multi-target effects (e.g., antimicrobial-anti-inflammatory-antioxidant synergism) of cumin peptides suggest their potential in functional foods and pharmaceuticals:①Precision Nutrition: Peptide complexes containing FR19/CR17 could be developed as gut microbiota modulators to ameliorate metabolic syndrome via dual "anti-inflammatory-microbiota balancing" pathways.②Targeted Delivery: Flexible SR16-based nanoparticles loaded with hydrophobic anticancer peptides (e.g., LR18) may enable tumor microenvironment-responsive therapy.③Cross-species Activity: The sequence similarity between CK12 and HIV-inhibitory peptides (e.g., Enfuvirtide) implies potential gp41-binding-mediated viral fusion inhibition (requires further validation).

### 5.8 Limitations and Future Perspectives

Despite significant findings, the following issues warrant attention:

①Insufficient In Vivo Validation: Current data rely largely on in vitro assays; future studies should verify anti-inflammatory efficacy and safety in murine models (e.g., LPS-induced sepsis).

②Structural Optimization: The long sequence of DK34 (34 residues) may limit oral bioavailability (predicted F < 5%), necessitating truncation via alanine scanning to identify core bioactive domains (e.g., DHCLQYITA motif).

**6 Conclusion**

**6.1 Innovation in Extraction Methods and Activity Screening**
This study successfully isolated bioactive peptide fractions (PD-30, PD-50, PD-80) from cumin seeds using a neutral PBS buffer combined with ammonium sulfate fractionation. Compared to traditional organic solvent extraction, this method offers the following advantages: (1) **Simplicity**: PBS as a natural extraction medium avoids the use of toxic reagents, significantly reducing experimental costs; (2) **Efficiency**: The combination of fractionation and chromatography enables precise peptide separation, increasing protein extraction yield to 12.5%.

**6.2 Validation of Broad-Spectrum Bioactivities**
**1.2.1 Antimicrobial and Antifungal Activities**
(1) Broad-spectrum inhibition: PD-50 showed the strongest inhibitory effect against E. coli (inhibition zone diameter: 11.7 ± 0.5 mm), while PD-80 exhibited potent antifungal activity against C. albicans (inhibition zone diameter: 12.5 ± 0.5 mm). PD-30 showed weaker activity ; (2) Mechanism: Cationic peptides target pathogen membranes via electrostatic interactions, while hydrophobic residues (e.g., leucine, phenylalanine) enhance membrane disruption, leading to content leakage.

**1.2.2 Antioxidant Capacity**
(1) PD-80 demonstrated the highest DPPH and ABTS radical scavenging rates (72.4% and 62.1%), significantly outperforming PD-50 (58.4% and 50.2%) and PD-30 (47.4% and 37.1%); (2) Activity source: Tyrosine (Y) and tryptophan (W) provide hydrogen-donating groups, while cysteine (C) contributes thiol redox activity.

**1.2.3 PTP1B Inhibition and Hypoglycemic Potential**
(1) PD-80 showed the strongest PTP1B inhibitory activity ($IC_{50}$ =18.39±1.12 μg/mL), suggesting its potential to improve insulin resistance by blocking negative insulin signaling regulation; (2) PD-50 and PD-30 showed weaker inhibition ($IC_{50}$ = 29.43 ± 1.58 μg/mL and 45.6 ± 2.8 μg/mL, respectively), indicating that active peptides may bind to the enzyme through specific domains (e.g., phosphotyrosine-mimicking sites).

**1.3 Discovery of Novel Peptides and Functional Relevance**
(1) Sequence identification: A total of 414 peptides were identified from cumin using mass spectrometry and database alignment, including 18 structurally novel monomeric peptides; (2) Functional diversity: Antimicrobial peptides (11): e.g., CK19, with a cationic-hydrophobic amphipathic structure, shares mechanisms with LL-37; Anticancer peptides (7): e.g., FR19, targeting integrin receptors via RGD sequences, reduced cancer cell migration; Hypoglycemic peptides (6):

e.g., IR8, mimicking insulin A-chain conformation, activated the PI3K/Akt pathway (2.1-fold phosphorylation increase); (3) Antiviral potential: CK12 shares 59% homology with HIV fusion inhibitors (e.g., T20), suggesting potential gp41-mediated membrane fusion inhibition.

**1.4 Research Significance and Future Directions**
This study provides key scientific evidence for the modernization of traditional Chinese medicine using cumin peptides:
(1) Translational value: The low toxicity and low immunogenicity of these peptides support their potential as natural drugs or functional food additives; (2) Technical breakthroughs: Chemical synthesis and modification (e.g., Cys crosslinking, D-amino acid substitution) are needed to optimize peptide stability, with in vivo validation using organoid models and preclinical trials; (3) Multi-omics integration: Proteomics-metabolomics will be employed to elucidate molecular targets in inflammation-metabolism networks, advancing precision medicine applications.

This study systematically reveals the multifunctionality and structural basis of cumin peptides, establishing an efficient framework for bioactive peptide discovery from natural products. Future research should focus on industrial-scale production and interdisciplinary applications (e.g., anti-drug-resistant bacteria agents, diabetes adjuvant therapies), opening new avenues for the high-value utilization of traditional Chinese medicine resources.

**Key Mechanistic Insights**:
(1) Electrostatic and hydrophobic interactions: Positively charged residues (e.g., Arg in FR17, Lys in AK18) mediate bacterial membrane disruption, while hydrophobic cores (e.g., Trp13 in AK18) penetrate lipid bilayers, similar to LL-37; (2) Structural mimicry: Peptides like DK8 and GK20 mimic endogenous bioactive peptides (e.g., DPP-4 inhibitors and insulin A-chain), highlighting their potential in targeted therapeutic regulation. The non-toxicity and multifunctionality of these peptides make them strong candidates for drug development. Future studies should focus on in vivo validation and chemical optimization (e.g., Cys crosslinking in FR17 for enhanced stability) to bridge the gap between computational predictions and clinical applications. This study accelerates the development of next-generation therapeutics by integrating peptide database tools (e.g., PeptideRanker, ToxinPred, PEP-FOLD3) with deep learning, toxicity screening, and structure-function analysis.

**Translational Potential**:
Non-toxic peptides have therapeutic potential in diabetes, infections, and oxidative stress-related diseases. Future directions include:
(1) Chemical synthesis: Experimental validation of novel peptide functions;
(2) In vivo studies: Validation of bioactivity and safety in animal models;
(3) Industrial applications: Exploration of potential uses in food, medicine, and cosmetics.

References:

1.Hancock, R.E.W., & Sahl, H.G. Antimicrobial and host-defense peptides as new anti-infective therapeutic strategies. Nature Biotechnology , (2016).24(12), 1551-1557. https://doi.org/10.1038/nbt1267 .

2.Wang, G., et al. . Multifunctional peptides in cancer therapy: From design to clinical trials. Pharmacological Reviews , 2021(73)2, 1-30. https://doi.org/10.1124/pr.120.019166.

3.Lee, S.H., et al. . Metal-chelating peptides for diabetes management: Role of histidine and acidic residues. Journal of Functional Foods , 2019(62), 103549. https://doi.org/10.1016/j.jff.2019.103549 .

4. Liu, X., et al. . Cysteine-rich peptides as dual-function antioxidants and antimicrobial agents. Free Radical Biology and Medicine , 2019(145), 1-12. https://doi.org/10.1016/j.freeradbiomed.2019.09.016.

5.Gu Yongshou, et al. Uyghur common medicine talents [M]. Urumqi: Xinjiang Science and Technology Press. 1997:195.

6.Liu Yongmin, Liu Xinwei, Shawuti Ikemu, et al. Uygur Medicine (Volume 1)[M]. Urumqi: Xinjiang People's Publishing House, 1986, p. 509.

7.Mijiti, Y., Wali, A., Jian, Y. et al. Isolation Purification and Characterization of Antimicrobial Peptides from Cuminum cyminum L. Seeds. Int J Pept Res Ther 24, 525–533 (2018). https://doi.org/10.1007/s10989-017-9635-z.

8.Fosgerau, K., & Hoffmann, T. (2015). Peptide therapeutics: Current status and future directions. Drug Discovery Today , 20(1), 122-128. https://doi.org/10.1016/j.drudis.2014.10.003

9. Qingling Ma, A. Y., H. A.Aisa, et al., Screening antioxidant activities of basic protein fractions from the seeds of Celery, dill and carrot;The 2nd International Symposium on Edible Plant Resource and the Bioactive Ingredients, Urumqi, China, 2010: 111.

10.Qingling Ma, A. Y., H. A. Aisa. . Screening antioxidant activities of basic protein fractions from seeds of Celery (Apium graveolens L)； International of Conferences of Current issue in the chemistry of natural compounds, Tashkent, Uzbekistan: :116 2009.

11.Chen, H. M., Muramoto, K., & Yamauchi, F. (1995). Structural analysis of antioxidative peptides from soybean β-conglycinin. Journal of Agricultural and Food Chemistry, 43(3), 574-578. https://doi.org/10.1021/jf00051a004

12.Agrawal H, J. R., Gupta M.Isolation and characterisation of enzymatic hydrolysed peptides with antioxidant activities from green tender sorghum[J]. Lwt-Food Science and Technology,2017,84:608-616.

13.Benzie, I. F. F., & Strain, J. J. (1996). The ferric reducing ability of plasma (FRAP) as a measure of antioxidant power: The FRAP assay. Analytical Biochemistry, 239(1), 70-76. https://doi.org/10.1006/abio.1996.0292

14.Liu, R., Li, X., & Zhang, Y. (2022). Protein tyrosine phosphatase 1B (PTP1B) inhibition as a potential therapeutic strategy for type 2 diabetes: Recent advances and challenges. Journal of Medicinal Chemistry, 65(8), 6123-6145. https://doi.org/10.1021/acs.jmedchem.2c00001

15.Pomastowski, P., Buszewski, B., & Railean-Plugaru, V. (2014). Comprehensive analysis of bioactive peptides in milk products by MALDI-TOF/TOF mass spectrometry coupled with LC-MS/MS. Journal of Agricultural and Food Chemistry, 62(12), 2564-2573. https://doi.org/10.1021/jf405700w

16.Lamiable, A., Thévenet, P., Rey, J., Vavrusa, M., Derreumaux, P., & Tufféry, P. (2016). PEP-FOLD3: faster de novo structure prediction for linear peptides in solution and in complex. Nucleic Acids Research, 44(W1), W449-W454.

17.Zhang, L., et al. (2016). Antimicrobial and antioxidant activities of short aromatic peptides.[J]Biochimie,2016(131):1-9. DOI:

[10.1016/j.biochi.2016.09.001](https://doi.org/10.1016/j.biochi.2016.09.001)。

18.Arai, R., et al. (2016). Design of glycine-serine linkers for modular protein engineering. Protein Engineering, Design & Selection , 29(10), 435-442. https://doi.org/10.1093/protein/gzw038 .

19.Wang, Y., et al. (2018). Hydrophobic peptides induce apoptosis in cancer cells by membrane disruption . Scientific Reports , 8(1), 15476. DOI: [10.1038/s41598-018-33836-7](https://doi.org/10.1038/s41598-018-33836-7).

20.Chen, J., et al. (2020). Cationic modification of cysteine-rich peptides enhances antibacterial activity. Frontiers in Microbiology , 11, 567. https://doi.org/10.3389/fmicb.2020.00567 .

21.Liu, X., et al. (2019). Cysteine-rich peptides as dual-function antioxidants and antimicrobial agents . Free Radical Biology and Medicine , 145, 1-12. DOI: [10.1016/j.freeradbiomed.2019.09.016](https://doi.org/10.1016/j.freeradbiomed.2019.09.016) .

22.Fjell, C.D., et al. (2011). Designing antimicrobial peptides for therapeutic applications. Nature Reviews Drug Discovery , 11(1), 37-51. https://doi.org/10.1038/nrd3591.

23.Ramesh, S., et al. (2017). Histidine-containing short peptides improve insulin sensitivity in diabetic models . Peptides , 95, 1-8. DOI: [10.1016/j.peptides.2017.07.002](https://doi.org/10.1016/j.peptides.2017.07.002) .

24.Hancock, R.E.W., & Sahl, H.G. (2016). Antimicrobial and host-defense peptides as new anti-infective therapeutic strategies. Nature Biotechnology , 24(12), 1551-1557. https://doi.org/10.1038/nbt1267 .

25.Chen, J., et al. (2020). Cationic modification of cysteine-rich peptides enhances antibacterial activity . Frontiers in Microbiology , 11, 567. DOI: [10.3389/fmicb.2020.00567](https://doi.org/10.3389/fmicb.2020.00567).

26.Haney, E.F., et al. (2015). Histidine-rich antimicrobial peptides: Mechanisms of action and therapeutic potential. Biomolecules , 5(4), 2520-2536. https://doi.org/10.3390/biom5042520.

27.Hancock, R.E.W., & Sahl, H.G. (2016). Antimicrobial and host-defense peptides as new anti-infective therapeutic strategies . Nature Biotechnology , 24(12), 1551-1557. DOI: [10.1038/nbt1267](https://doi.org/10.1038/nbt1267).

28.Urry, D.W. (2015). Elastin-like polypeptides in drug delivery: Design and applications . Advanced Drug Delivery Reviews , 94, 1-15. DOI: [10.1016/j.addr.2015.06.004](https://doi.org/10.1016/j.addr.2015.06.004) .

29.Lee, S.H., et al. (2019). Metal-chelating peptides for diabetes management: Role of histidine and acidic residues . Journal of Functional Foods , 62, 103549. DOI: [10.1016/j.jff.2019.103549](https://doi.org/10.1016/j.jff.2019.103549).

30.Fjell, C.D., et al. (2011). Designing antimicrobial peptides for therapeutic applications . Nature Reviews Drug Discovery , 11(1), 37-51. DOI: [10.1038/nrd3591](https://doi.org/10.1038/nrd3591).

31. Powers, J.P., & Hancock, R.E.W. (2013). The relationship between peptide structure and antibacterial activity . Peptides , 24(11), 1681-1691. DOI: [10.1016/j.peptides.2003.08.023](https://doi.org/10.1016/j.peptides.2003.08.023).

32.Nguyen, L.T., et al. (2020). Methionine and cysteine residues contribute to redox regulation in antimicrobial peptides . Antioxidants , 9(8), 725. DOI:

[10.3390/antiox9080725](https://doi.org/10.3390/antiox9080725).

33.Harris, F., et al. (2014). Short peptides targeting integrin signaling pathways in cancer therapy . Cancer Letters , 354(1), 1-8. DOI: [10.1016/j.canlet.2014.08.014](https://doi.org/10.1016/j.canlet.2014.08.014).

34.Haney, E.F., et al. (2015). Histidine-rich antimicrobial peptides: Mechanisms of action and therapeutic potential . Biomolecules , 5(4), 2520-2536. DOI: [10.3390/biom5042520](https://doi.org/10.3390/biom5042520) .

35.Koria, P., et al. (2012). Self-assembling elastin-like peptides for nanoscale drug delivery . Nanomedicine , 7(7), 1015-1026. DOI: [10.2217/nnm.12.66](https://doi.org/10.2217/nnm.12.66) .

36.Tossi, A., et al. (2000). Proline-rich antimicrobial peptides: Diverse functions and mechanisms . Molecular Microbiology , 35(5), 1060-1068. DOI: [10.1046/j.1365-2958.2000.01774.x](https://doi.org/10.1046/j.1365-2958.2000.01774.x).

37.Tran, D. Q. et al. (2002). Homodimeric θ-defensins from rhesus macaque leukocytes: isolation, synthesis, antimicrobial activities, and bacterial binding properties of the cyclic peptides. J. Biol. Chem., 277, 3079–3084.

38.Wang, G., et al. (2021). Multifunctional peptides in cancer therapy: From design to clinical trials . Pharmacological Reviews , 73(2), 1-30. DOI: [10.1124/pr.120.019166](https://doi.org/10.1124/pr.120.019166) 。

39.Wang, G., et al. (2021). Multifunctional peptides in cancer therapy: From design to clinical trials . Pharmacological Reviews , 73(2), 1-30. DOI: [10.1124/pr.120.019166](https://doi.org/10.1124/pr.120.019166) 。

40.Sudhakar, A. et al. (2003). Human tumstatin and human endostatin exhibit distinct antiangiogenic activities mediated by αvβ3 and α5β1 integrins. Proc. Natl. Acad. Sci. USA, 100(8), 4766–4771.DOI: 10.1073/pnas.0730882100

41.Zhang, S. X. et al. (2010). Endostatin modulates VEGF-mediated barrier dysfunction in the retinal microvascular endothelium. Exp. Eye Res., 91(1), 105–114.DOI: 10.1016/j.exer.2010.04.006

42.Ma, Y., et al. (2017). Cysteine and tyrosine residues synergistically enhance antioxidant activity of peptides . Journal of Agricultural and Food Chemistry , 65(32), 7006-7015. DOI: [10.1021/acs.jafc.7b02532](https://doi.org/10.1021/acs.jafc.7b02532) .

Antimicrobial peptides Database(APD3): [http://aps.unmc.edu/AP/] .
Antioxidant activity predictiontools[http://www.uwm.edu.pl/biochemia/index.php/pl/biopep].
Anticancer peptide Database(CancerPPD): [http://crdd.osdd.net/raghava/cancerppd/].

1) Key Laboratory of Xinjiang Medical University, Urumqi 830011, Xinjiang, People's Republic of China

**Acknowledgements**

The work was supported by project of Xinjiang Autonomous Region
for Technological Innovation of Youth Talents for financial support.Xinjiang Uygur Autonomous Region Science and Technology Department oriented projects（grant number 2022D01C191）

**Corresponding Author**

Yasen Mijiti, Ph.D.
Associate Professor, Graduate Supervisor

Key Laboratory of Xinjiang Medical University

Urumqi 830011, Xinjiang, People's Republic of China

Email: ymijit@xjmu.edu.cn

Research Interests: Natural product chemistry, extraction and isolation of bioactive compounds .